\UseRawInputEncoding
\documentclass[
 article,
superscriptaddress,
twocolumn,
 amsmath,amssymb,
 aps,
prb,
]{revtex4-2}

\usepackage{graphicx}% Include figure files
\usepackage{dcolumn}% Align table columns on decimal point
\usepackage{bm}% bold math
\usepackage{tikz}
\usepackage{mathtools}
\usepackage[caption=false]{subfig}
\usepackage{comment}
\usepackage{xcolor}
\begin{document}

\preprint{APS/123-QED}

\title{Switchable Surface Linear Photogalvanic Effect in the Magnetic Weyl Semimetal Co$_3$Sn$_2$S$_2$}

\author{Niket Shah}
\affiliation{Department of Physics, Indian Institute of Technology Bombay, Powai, Mumbai 400076, India}
\author{Aymen Nomani}
\affiliation{Department of Physics and Texas Center for Superconductivity, University of Houston, Houston, TX 77204, USA}
\author{Kai Chen}
\affiliation{School of Physics Science and Engineering, Tongji University, 200092 Shanghai, China}
\author{Hridis K. Pal}
\affiliation{Department of Physics, Indian Institute of Technology Bombay, Powai, Mumbai 400076, India}
\author{Pavan Hosur}
\affiliation{Department of Physics and Texas Center for Superconductivity, University of Houston, Houston, TX 77204, USA}

\date{\today}

\begin{abstract}

We investigate the linear photogalvanic effect (LPGE) on the surface of the magnetic Weyl semimetal Co$_3$Sn$_2$S$_2$ using a Green's-function and diagrammatic formalism. While the LPGE vanishes in the centrosymmetric bulk, it is symmetry-allowed on the surface where inversion symmetry is broken. We show that unitary crystal symmetries on the surface produce characteristic sign reversals of the total photocurrent at certain polarization angles upon flipping the magnetization. We further find that the intrinsic contribution to the LPGE is strongly constrained by an antiunitary mirror symmetry, which forces several nonlinear response tensor elements to vanish. In contrast, the extrinsic contribution is not subject to these constraints and displays a large magnitude which, we argue, is due to the enhanced density of states associated with Fermi-arc surface states. The current exhibits an approximately linear temperature dependence and a low-frequency power-law scaling, $|j_y|\propto \omega^{-2.2}$, with weak temperature dependence of the scaling exponent. Our results identify $\mathrm{Co}_3\mathrm{Sn}_2\mathrm{S}_2$ as a promising platform for experimentally accessing symmetry-controlled nonlinear transport in realistic systems and for applications in magnetically controlled optoelectronic devices.

\end{abstract}

\maketitle

\section{Introduction}

Weyl semimetals (WSMs) belong to a broader class of three-dimensional (3D) topological quantum materials whose electronic properties are governed by topology and symmetry. In these systems the bulk bands are nondegenerate and intersect only at discrete points in momentum space known as Weyl nodes, around which the low-energy quasiparticles obey a Weyl Hamiltonian \cite{VafekDiracReview,Burkov2018,Burkov:2016aa,YanFelserReview,Armitage2018,Shen2017,Belopolski:2016wu,Guo2018,Chang2016,Gyenis_2016,Huang:2015vn,Inoue1184,Lv:2015aa,Sun2015a,Xu2015,Xu2016,Xu613,Yang:2015aa,Zheng2016}. The nondegeneracy condition requires breaking either inversion symmetry, time-reversal symmetry, or both, since the simultaneous presence of these symmetries enforces doubly degenerate bands. Each Weyl node acts as a monopole of Berry curvature and carries a quantized chirality, implying that Weyl nodes occur in pairs of opposite chirality and are topologically protected against gap opening unless annihilated pairwise.

A hallmark feature of WSMs is the appearance of topological surface states known as Fermi arcs (FAs), which form open contours connecting the surface projections of Weyl nodes of opposite chirality in the surface Brillouin zone \cite{Hosur2013a,Wang_2018,Hu:2019aa,ZyuninBurkovWeylTheta,ChenAxionResponse,VazifehEMResponse,Burkov_2015,Hosur2012,Wang2017,Halterman2018,Halterman2019,Nagaosa:2020aa,NielsenABJ,IsachenkovCME,SadofyevChiralHydroNotes,Loganayagam2012,GoswamiFieldTheory,Wang2013,BasarTriangleAnomaly,LandsteinerAnomaly}. These states reflect the topological connection between bulk and surface electronic structures, as their wavefunctions penetrate the bulk and connect opposite surfaces. 

Despite extensive work on bulk responses in WSMs, comparatively less attention has been paid to phenomena dominated by FA surface states. Yet many experimental probes are inherently surface sensitive and therefore naturally suited to accessing responses arising from these states. Studying such responses provides a direct route to isolating surface contributions and probing the physics of FAs. Some progress has been made in understanding surface physics, including studies of the surface dc conductivity \cite{Pal2022} and surface superconductivity \cite{Nomani2023IntrinsicSemimetals,KuibarovNature2024,SchimmelNatComm2024,XingNSR2019,Moreno2026}. Recent theoretical and experimental works have shown that confinement-induced photogalvanic responses are intimately linked to the connectivity and geometry of the FAs \cite{Steiner2022}, and that crystal symmetries can be exploited to disentangle surface and bulk photocurrents \cite{ChangPRL2020,ReesPRL2021}. Together, these developments reveal a rich surface state phenomenology that can be accessed through surface-dominated responses and motivate this work.

Nonlinear optical effects provide a powerful probe of topological electronic structure because they are highly sensitive to symmetry and Berry curvature physics. In systems lacking inversion symmetry, second-order responses such as second harmonic generation, the linear photogalvanic effect (LPGE), and the circular photogalvanic effect (CPGE) can generate dc photocurrents under optical illumination \cite{Manuel2025,Kolesnikov2019,Vijaysankar2026,Zyuzin2018,Morimoto2016,Steiner2022,Cao2022,Zhurun2019,Li2018,Gao2021}. In WSMs, these responses can acquire distinctive topological signatures; in particular, the CPGE can exhibit a quantized photocurrent generation rate determined by the Weyl node chirality and fundamental constants \cite{Juan2017,QuantCPGE,CPGETaAs}. Experiments have observed nonlinear optical responses in several nonmagnetic noncentrosymmetric WSMs, including $\beta$-WP$_2$ \cite{Lv2021}, RhSi \cite{Lu2022}, $\mathrm{Cd}_3\mathrm{As}_2$ \cite{Liang2022}, and TaAs \cite{Patankar2018,Ma2017a,Osterhoudt2019}.

Magnetic WSMs provide a particularly attractive setting for nonlinear optical phenomena because intrinsic magnetization provides an additional tunable parameter that can modify the Weyl node configuration and associated response functions. Among these materials, the kagome ferromagnet $\mathrm{Co}_3\mathrm{Sn}_2\mathrm{S}_2$ has emerged as a prominent example. This compound hosts well-separated Weyl nodes and exhibits a variety of striking phenomena, including a giant anomalous Hall effect, a large anomalous Nernst effect, and a strong magneto-optical response \cite{Liu2018a,Okamura:2020tq,PhysRevMaterials.4.024202}. Many of these responses depend sensitively on the direction of magnetization and other external parameters \cite{Ozawa2024,JinyingYang,ShubkovDeHaas2026}, making $\mathrm{Co}_3\mathrm{Sn}_2\mathrm{S}_2$ a promising platform for spintronic, optoelectronic, and thermoelectric applications.

This work explores the surface LPGE in $\mathrm{Co}_3\mathrm{Sn}_2\mathrm{S}_2$. This effect emerges from the interplay of topology, symmetry, and magnetism. The WSM phase supports FA surface states with a substantial low-energy density of states, while the bulk centrosymmetry suppresses bulk photogalvanic effects and the intrinsic ferromagnetic order provides a controllable magnetization degree of freedom that can be used to manipulate the resulting photocurrent. Motivated by this interplay, we investigate the LPGE on the surface of $\mathrm{Co}_3\mathrm{Sn}_2\mathrm{S}_2$ using a tight-binding model of its electronic structure. We systematically analyze the temperature and frequency dependence of the resulting photocurrent and find the response to be easily within experimental reach. The large magnitude of the effect is consistent with the finite surface density of states associated with the FAs and contrasts with photogalvanic responses associated with isolated surface Dirac cones in topological insulators~\cite{Hosur2011a}. This interpretation is further supported by the approximately linear temperature dependence of the LPGE, indicating that the response is dominated by low-energy excitations near the Fermi arc.

A symmetry analysis shows that an antiunitary mirror symmetry, given by time reversal followed by reflection, constrains the intrinsic LPGE by eliminating tensor components that are even under magnetization reversal for out-of-plane magnetization. As a result, the intrinsic contribution is purely odd under inversion of the magnetization. In contrast, extrinsic contributions arising from scattering processes are not subject to this constraint and can generate both even- and odd-in-magnetization components. However, the angular dependence of the LPGE provides a means to isolate the magnetization-odd response. For a given current direction, specific polarization angles suppress the symmetry-allowed even-in-magnetization contributions, leaving a response that is strictly odd under magnetization reversal. This enables a controlled inversion of the photocurrent with magnetization, providing a route to magnetically tunable optoelectronic functionality.

This paper is organized as follows. Sec. \ref{pge} introduces the photogalvanic effect and summarizes the procedure followed to compute it in this work. In Sec. \ref{sym}, the symmetries of $\text{Co}_3\text{Sn}_2\text{S}_2$ and their dependence on magnetization direction are discussed. Finally, in Sec. \ref{num}, the effects of the symmetries of $\text{Co}_3\text{Sn}_2\text{S}_2$ on its surface photogalvanic effect are discussed together with the details of the numerical studies and the results of the study are presented.

\section{Surface Photogalvanic Effect}
\label{pge}

In this section, we describe the calculation of the photogalvanic effect for the magnetic WSM Co$_3$Sn$_2$S$_2$. 

Consider oscillating electric fields propagating in the $z$-direction. The non-linear second-order current response to two electric fields oscillating at $\omega$ and $\omega'$, $\vec{E}_\omega  e^{i (k_z z - \omega t)}$ and $\vec{E}_{\omega'}  e^{i (k_z z - \omega' t)}$ is defined by
\begin{equation}
j_{\omega+\omega'}^{\mu}	=\sum_{\alpha\beta } \chi^{\mu\alpha\beta}_{\omega,\omega'}E_{\omega}^{\alpha}E_{\omega'}^{\beta}    
\end{equation}
where $\chi^{\mu\alpha\beta}_{\omega,\omega'}$ is the second-order response function given by a rank-three tensor. When $\omega'=-\omega$, the response is a dc response known as the photogalvanic effect. This is the response we will focus on in the rest of this paper.

The presence of symmetries constrains the allowed values of  $\chi^{\mu\alpha\beta}$. To start with, for  $j_{0}^{\mu}$ to be real for a real electric field, $\chi^{\mu\alpha\beta}=\chi^{\mu\beta\alpha^{*}}$, which implies that the real (imaginary) part of $\chi^{\mu\alpha\beta}$ is symmetric (anti-symmetric) under $\alpha\longleftrightarrow\beta$. Henceforth, the 0 subscript on $j^\alpha$ and the subscripts $\omega,-\omega$ on $\chi^{\mu\alpha\beta}$ will be dropped to avoid notational clutter. Let us write the current as the sum of symmetric and anti-symmetric parts:
\begin{align} 
j^{\mu} =S^{\mu\alpha\beta}\frac{E_{\omega}^{\alpha}E_{-\omega}^{\beta}+E_{\omega}^{\beta}E_{-\omega}^{\alpha}}{2}+A^{\mu\alpha\beta}\frac{E_{\omega}^{\alpha}E_{-\omega}^{\beta}-E_{\omega}^{\beta}E_{-\omega}^{\alpha}}{2} 
\end{align}
where $S^{\mu\alpha\beta}$ $(A^{\mu\alpha\beta})$ is the symmetric (antisymmetric) part of $\chi^{\mu\alpha\beta}$. The CPGE is captured by $A^{\mu\alpha\beta}$, the LPGE corresponds to the traceless part of $S^{\mu\alpha\beta}$ when viewed as a matrix for fixed $\mu$, while $S^{\mu\alpha\alpha}$ defines the response to unpolarized light. 

\begin{figure}[h] 
\includegraphics[width=1.0\columnwidth]{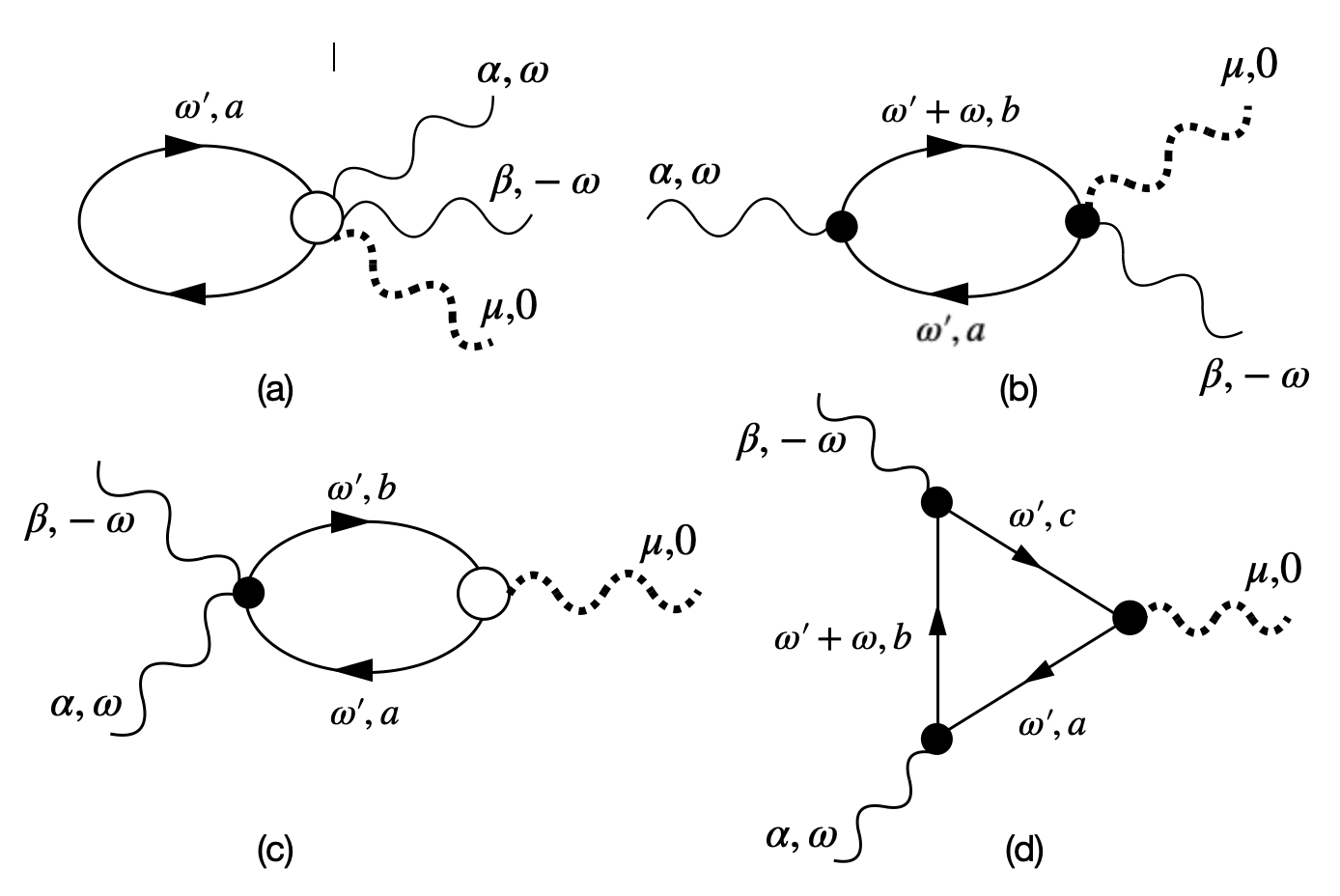}
\caption{\label{fig:fdiagram}  
 The four Feynman diagrams used to calculate $\chi^{\mu \alpha \beta }$. The dotted lines represent the output current $j$ while the wavy solid lines represent external fields.}
\end{figure}

In general, any second-order response $\chi^{\mu\alpha\beta}$ can be calculated using four Feynman diagrams, as shown in Fig. \ref{fig:fdiagram}.
Defining $\hat{H}_{\boldsymbol{k}}$ as the Bloch Hamiltonian for a WSM slab consisting of $L$ atomic layers, where $\boldsymbol{k}$ is the 2D momentum parallel to the surface, the mathematical expressions corresponding to the four diagrams are \cite{Diagrammatic}

\begin{align}
\label{1000}
&\chi^{\mu\alpha\beta} = \frac{e^{3}}{(\hbar\omega)^{2}} \sum_{a,b,c} \intop_{\boldsymbol{k}} \left[ \frac{1}{2}f_{a}h^{\mu\alpha\beta}\right.  \notag\\
& + f_{ab}\left(\frac{h^{\alpha}_{ab}h^{\mu\beta}_{ba}}{\omega-\varepsilon_{ab}+i\delta} - \frac{1}{2}\frac{h^{\alpha\beta}_{ab}h^{\mu}_{ba}}{\varepsilon_{ab}+i\delta}\right)  \notag\\
&  - h^{\alpha}_{ab}h^{\beta}_{bc}h^{\mu}_{ca} \frac{(\omega-\varepsilon_{bc})f_{ab} + (\omega-\varepsilon_{ba})f_{cb}}{(\omega-\varepsilon_{ba}+i\delta)(\omega-\varepsilon_{cb}+i\delta)(\varepsilon_{ca}+i\delta)}\notag \\
&+ \big[(\alpha, \omega) \leftrightarrow (\beta, -\omega)\big]\bigg].
\end{align}
$\chi^{\mu\alpha\beta}$ is expressed in terms of band indices $a$, $b$, and $c$, where the integration is performed over the Brillouin zone. The quantity
%$\Delta k = \frac{\text{Brillouin zone area}}{\text{Number of k-points} }$ and 
$\hat{h}^{\alpha_1 .... \alpha_n }=\frac{\partial}{\partial k_{\alpha_1}}....\frac{\partial}{\partial k_{\alpha_n}}[ \hat{H}_{\boldsymbol{k}}]$ is the $n^{th}$ derivative of the Bloch Hamiltonian with respect to crystal momentum. $f_n$ is the Fermi-Dirac distribution corresponding to band $n$, and $f_{ab} = f_a - f_b$. The band energy differences are defined as $\varepsilon_{ab} = \varepsilon_a - \varepsilon_b$, where $\varepsilon_n$ is the energy of band $n$. The parameter $\delta$ is a small positive number that accounts for a finite quasiparticle lifetime due to scattering, with $\delta \sim \hbar/(2\tau)$, where $\tau$ is the scattering time.

In the present approach, the bulk Bloch Hamiltonian is used to evaluate the band energies and their momentum derivatives. As discussed earlier, the bulk Hamiltonian preserves inversion symmetry, which enforces the vanishing of the bulk photogalvanic response. Consequently, any finite contribution to $\chi^{\mu\alpha\beta}$ must originate from inversion symmetry breaking at the surface. Eq.~(\ref{1000}) therefore yields the photogalvanic conductivity from the surfaces of the slab.

\section{$\text{Co}_3\text{Sn}_2\text{S}_2$ Symmetries}
\label{sym}
\begin{figure}[htbp]
    \centering

    \subfloat[]{
        \includegraphics[width=0.48\columnwidth]{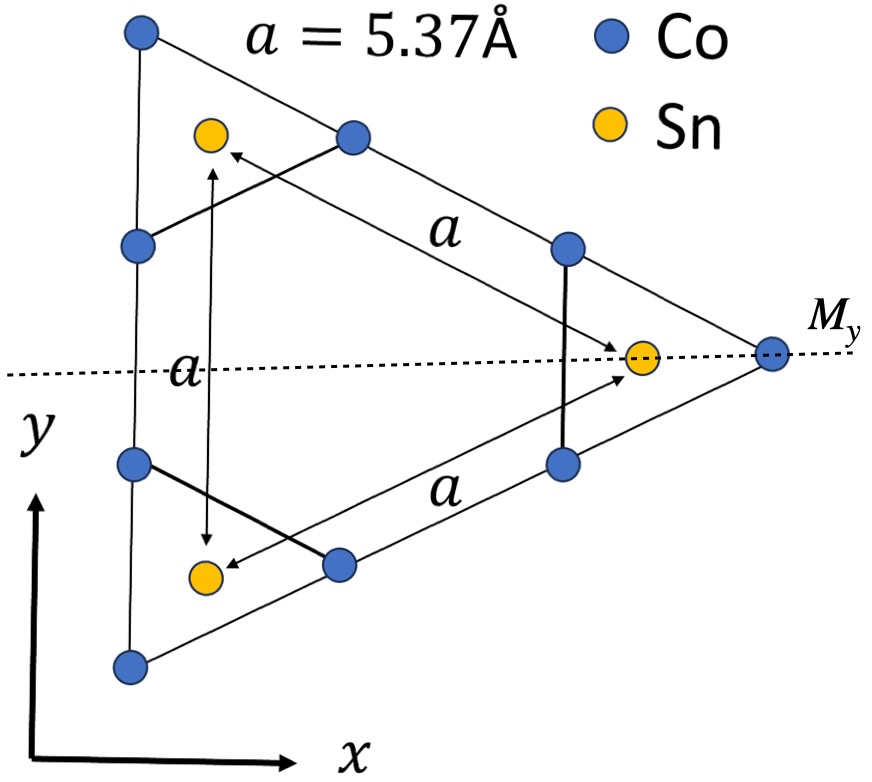}
        \label{fig:structure1}
    }
    \subfloat[]{
        \includegraphics[width=0.48\columnwidth]{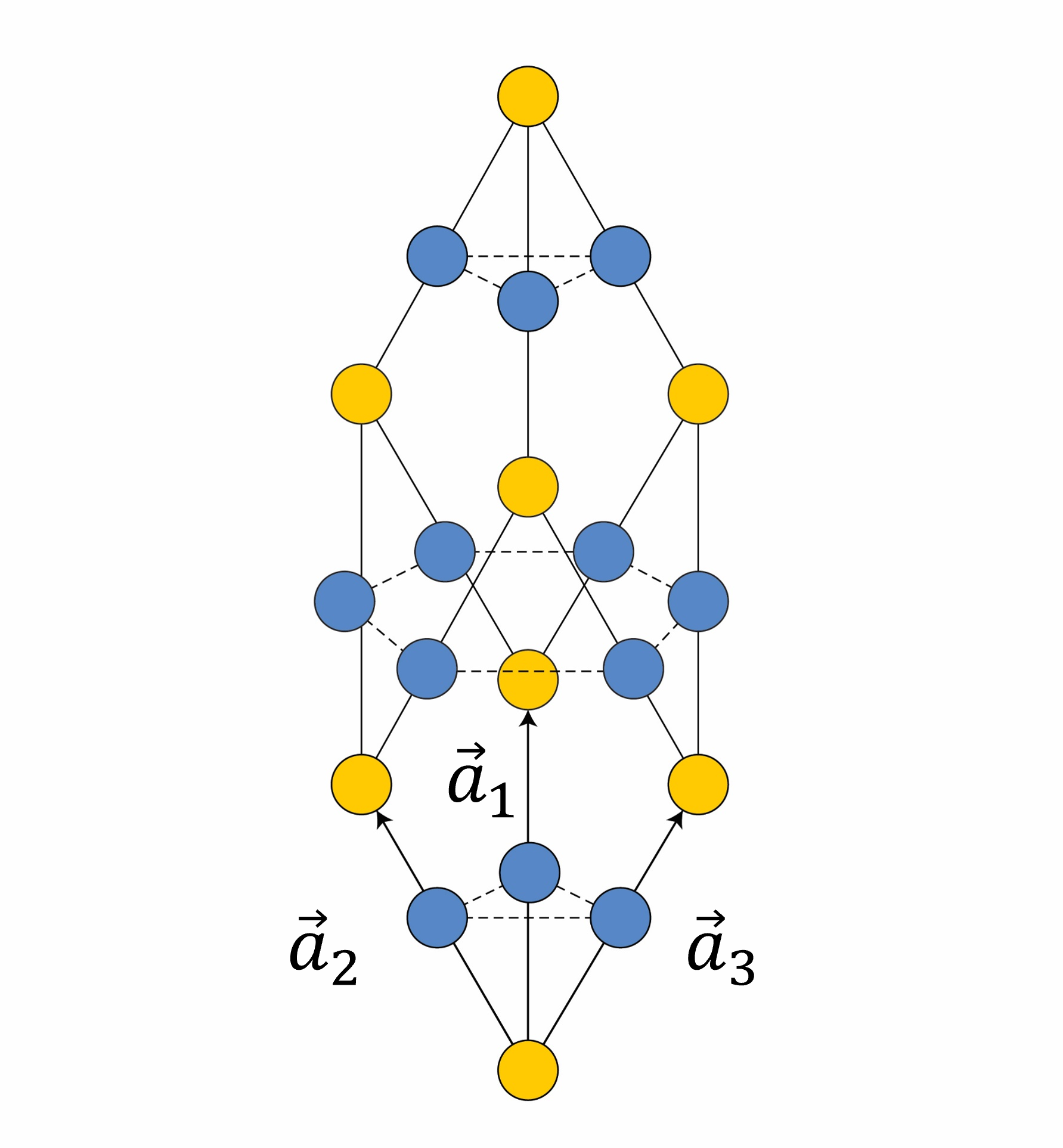}
        \label{fig:structure2}
    }
    
    \subfloat[]{
        \includegraphics[width=0.8\columnwidth]{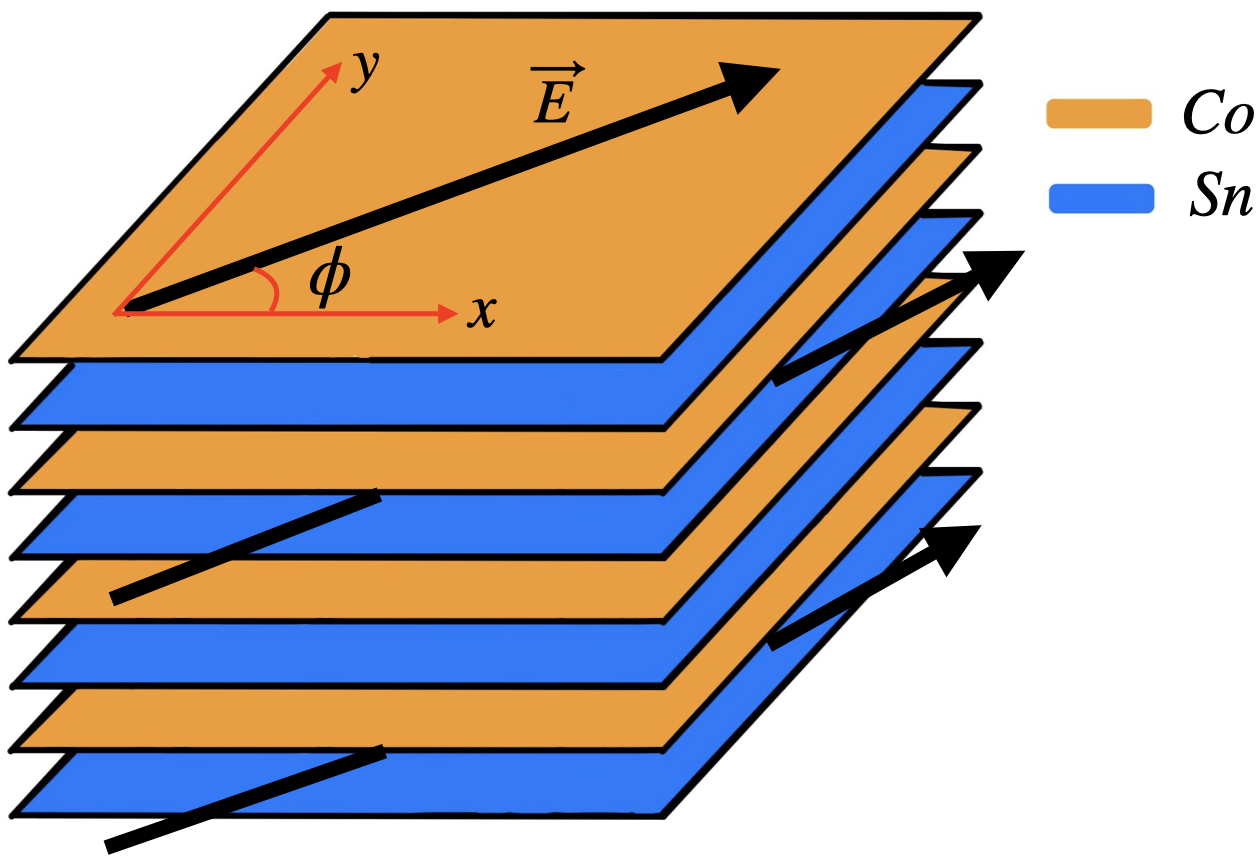}
        \label{fig:pdiagram} 
    }
    
    \caption{
    Effective structure of $\text{Co}_3\text{Sn}_2\text{S}_2$ with the S-layer and the Sn-atoms in the Co layer ignored along with the experimental geometry.
    % spectral functions of Co and Sn layers.
    (a) Kagome structure of the Co layer and the position of the triangular Sn layer relative to the Co layer in the $x$-$y$ plane. (b) Tetrahedral unit cell formed by the Sn atoms and the lattice translation vectors $\vec{a}_i$.
    % (c) and (d) Spectral functions $A(\boldsymbol{k},0)$ of the Co and Sn surfaces. The dotted hexagon indicates the first Brillouin Zone. The Co surface has a higher density of states compared to the Sn surface.
    (c) Schematic showing the layered structure of  $\text{Co}_3\text{Sn}_2\text{S}_2$ with a uniform ac electric field $\vec{E}$ applied to the slab at polarization angle $\phi$.
    }
    \label{fig:structure}
\end{figure}

$\text{Co}_3\text{Sn}_2\text{S}_2$ has a rhombohedral lattice structure (Fig. \ref{fig:structure2})  with $R\Bar{3}m$ space group \cite{Schnelle2013}, which consists of a Kagome layer of Co atoms with an Sn atom in the middle of the hexagon, a triangle layer of Sn atoms and a triangle layer of S atoms. The corners of the triangles formed in the Co-layer Kagome lattice are sandwiched on one side in the  $z$-direction by an Sn-layer Sn atom and on the other side by an S-layer S atom. Each successive Sn-layer gets translated in the primitive translation vector directions, such that every 4th layer comes back on top of the 1st layer exactly. The resulting structure looks like a rhombohedral with the Sn-layer Sn atoms forming the corners of the rhombohedral.

The $R\bar{3}m$ crystal structure possesses inversion symmetry, threefold rotational symmetry $C_{3z}$, and mirror symmetries with respect to the $x$--$z$ plane ($M_y$) and its $C_{3z}$-related counterparts. In the absence of magnetization, the system additionally preserves time-reversal symmetry $\mathcal{T}$ and spin-rotation symmetry about the $z$ axis ($\sigma_z$). The presence of inversion symmetry prohibits any bulk photogalvanic response; however, this symmetry is explicitly broken at the surface, thereby allowing a finite photogalvanic effect.

$\text{Co}_3\text{Sn}_2\text{S}_2$ is a ferromagnet with strong magnetic anisotropy and a Curie temperature of approximately $175\,\mathrm{K}$ ($\sim 0.015\,\mathrm{eV}$) \cite{Liu2018a}. The emergence of magnetization, whether intrinsic or induced by an external magnetic field, breaks time-reversal symmetry $\mathcal{T}$. When the magnetization lies in the $x$--$z$ plane, the mirror symmetry $M_y$ is broken; however, the combined anti-unitary symmetry $\mathcal{T}\otimes M_y$ remains preserved. A summary of the transformation properties of relevant physical quantities under these symmetry operations is provided in Table~\ref{tab:sym}.

\begin{table}
\caption{\label{tab:sym} Transformation properties of relevant quantities under symmetry operators}
\begin{ruledtabular}
\begin{tabular}{cccccccc}
 &$I$ & $M_y$& $\mathcal{T}$ & $\mathcal{T}\otimes M_y $\\
\hline
$E_x$& $-E_x$ &  $E_x$  &   $E_x$ & $E_x$ \\
$E_y$& $-E_y$ &  $-E_y$ &   $E_y$ & $-E_y$      \\
$j_x$& $-j_x$ &  $j_x$  &  $-j_x$ & $-j_x$ \\
$j_y$& $-j_y$ &  $-j_y$ &  $-j_y$ & $j_y$\\
$m_z$& $m_z$  &  $-m_z$ &  $-m_z$ & $m_z$\\
$m_x$& $m_x$  &  $-m_x$ &  $-m_x$ & $m_x$ \\
$m_y$& $m_y$  &  $m_y $ &  $-m_y$ & $-m_y$\\

\end{tabular}
\end{ruledtabular}

\end{table}

\begin{table}
\caption{\label{tab:zero} $\chi^{\mu\alpha\beta}$ terms (intrinsic contributions) constrained to 0 for magnetization in the three principal directions and the symmetries which enforce them. }
\begin{ruledtabular}
\begin{tabular}{cccccccc}
 &z & $y$& x\\
\hline
$\chi^{xxx}$&  $\mathcal{T}\otimes M_y$       &           & $\mathcal{T}\otimes M_y$  \\
$\chi^{xxy}$&                         &   $M_y$  &                 \\
$\chi^{xyx}$&                         &   $M_y$  &                   \\
$\chi^{xyy}$&  $\mathcal{T}\otimes M_y$       &            & $\mathcal{T}\otimes M_y$ \\
$\chi^{yxx}$&                         &   $M_y$  &                 \\
$\chi^{yxy}$&  $\mathcal{T}\otimes M_y$       &            & $\mathcal{T}\otimes M_y$ \\
$\chi^{yyx}$&  $\mathcal{T}\otimes M_y$      &            & $\mathcal{T}\otimes M_y$ \\
$\chi^{yyy}$&                         &   $M_y$                      \\

\end{tabular}
\end{ruledtabular}

\end{table}

The second-order conductivity tensor $\chi^{\mu\alpha\beta}$ contains both intrinsic and extrinsic contributions. The intrinsic contribution is determined by the band structure and is therefore strongly constrained by crystal symmetries. In contrast, the extrinsic contribution depends explicitly on the quasiparticle lifetime through a finite imaginary self-energy. As a result, certain symmetry constraints derived below apply specifically to the intrinsic response, while others constrain the full conductivity tensor including both intrinsic and extrinsic contributions. In the present system, the extrinsic contribution dominates the magnitude of the photocurrent.

For a general magnetization direction $\mathbf{m}=(m_x,m_y,m_z)$, the intrinsic nonlinear conductivity tensor transforms under the mirror symmetry $M_y$ and the antiunitary mirror symmetry $\mathcal{T}M_y$ as
\begin{align}
M_y:\quad
\chi^{\mu\alpha\beta}_{m_xm_ym_z}
&\rightarrow
(-1)^{N_y}
\chi^{\mu\alpha\beta}_{-m_x\,m_y\,-m_z},
\label{1004}
\\
\mathcal{T}M_y:\quad
\chi^{\mu\alpha\beta}_{m_xm_ym_z}
&\rightarrow
(-1)^{N_y+1}
\chi^{\mu\alpha\beta}_{m_x\,-m_y\,m_z},
\label{1002}
\end{align}
where $N_y$ denotes the number of $y$ indices among $\mu$, $\alpha$, and $\beta$. The extrinsic contributions obey \eqref{1004} but not \eqref{1002} as the latter is antiunitary and reverses the sign of the broadening term $\delta$ in \eqref{1000}. These relations impose general selection rules on the intrinsic LPGE tensor. For magnetization along $\hat y$, Eq.~\eqref{1004} permits only tensor components with even $N_y$ for both intrinsic and extrinsic responses. In contrast, for magnetization along $\hat x$ or $\hat z$, Eq.~\eqref{1002} permits only tensor components with odd $N_y$ for intrinsic response but does not constrain the extrinsic response. Thus, the antiunitary mirror symmetry $\mathcal{T}M_y$ removes the even-$N_y$ intrinsic tensor structure whenever the magnetization lies in the $xz$ plane.

We now specialize to the experimentally relevant case $\mathbf m\parallel \pm\hat z$. In this configuration, the threefold rotational symmetry $C_{3z}$ constrains the full in-plane second-order conductivity tensor, including both intrinsic and extrinsic contributions, to the form \cite{HuaJiang2025}
\begin{align}
\chi^{xxx}
&=
-\chi^{xyy}
=
-\chi^{yxy}
=
-\chi^{yyx},
\label{1001}
\\
\chi^{yyy}
&=
-\chi^{yxx}
=
-\chi^{xxy}
=
-\chi^{xyx}.
\label{1003}
\end{align}
The tensor components in Eq.~\eqref{1001} contain an even number of $y$ indices, while those in Eq.~\eqref{1003} contain an odd number of $y$ indices. Consequently, for $\mathbf m\parallel \hat z$, the intrinsic contribution is restricted entirely to the tensor structure in Eq.~\eqref{1003}$,$ while Eq.~\eqref{1001} contributes only through extrinsic processes.

The linear and circular photogalvanic currents generated by an in-plane optical field are
\begin{align}
j_L^\mu
=
|E|^2
\Big[
&
\chi^{\mu xx}\cos^2\phi
+
\chi^{\mu yy}\sin^2\phi
\nonumber\\
&
+
\left(
\chi^{\mu xy}
+
\chi^{\mu yx}
\right)
\sin\phi\cos\phi
\Big],
\end{align}
where $\phi$ is the polarization angle of the incident linearly polarized field, and
\begin{align}
j_C^\mu
=
\pm
\frac{|E|^2}{2}
\left(
\chi^{\mu xy}
-
\chi^{\mu yx}
\right),
\end{align}
where the two signs correspond to opposite circular polarizations. Equations~\eqref{1001} and \eqref{1003} imply $\chi^{\mu xy}=\chi^{\mu yx}$, and therefore the CPGE vanishes identically for both intrinsic and extrinsic contributions. The LPGE currents can be written as
\begin{align}
j_L^x
&=
|E|^2
\left[
\chi^{xxx}\cos 2\phi
+
\chi^{xyx}\sin 2\phi
\right],
\\
j_L^y
&=
|E|^2
\left[
\chi^{yxx}\cos 2\phi
-
\chi^{xxx}\sin 2\phi
\right].
\end{align}

The intrinsic contribution to $\chi^{xxx}$ vanishes due to $\mathcal{T}M_y$, so the intrinsic LPGE for $\mathbf m\parallel \hat z$ is entirely determined by the odd-$N_y$ tensor structure in Eq.~\eqref{1003} and is therefore odd under $m_z\rightarrow -m_z$ at all polarization angles. The extrinsic response, however, can generate both even- and odd-in-$m_z$ contributions through the tensor structure in Eq.~\eqref{1001}. Nevertheless, the odd-in-$m_z$ response can still be isolated at specific polarization angles. For example,
\begin{align}
j_L^x(\phi=\pi/4)
&=
|E|^2 \chi^{xyx},
\label{eq:jx-odd}\\
j_L^y(\phi=0)
&=
|E|^2 \chi^{yxx}\label{eq:jy-odd},
\end{align}
depend only on odd-$N_y$ tensor components and therefore reverse sign under $m_z\rightarrow -m_z$. These experimentally accessible configurations isolate the magnetization-odd LPGE response while retaining both intrinsic and extrinsic contributions.

In the absence of magnetization, both $\mathcal T$ and $M_y$ symmetries are simultaneously preserved, forcing the entire second-order conductivity tensor to vanish.

\begin{figure*} 
\includegraphics[width=2\columnwidth]{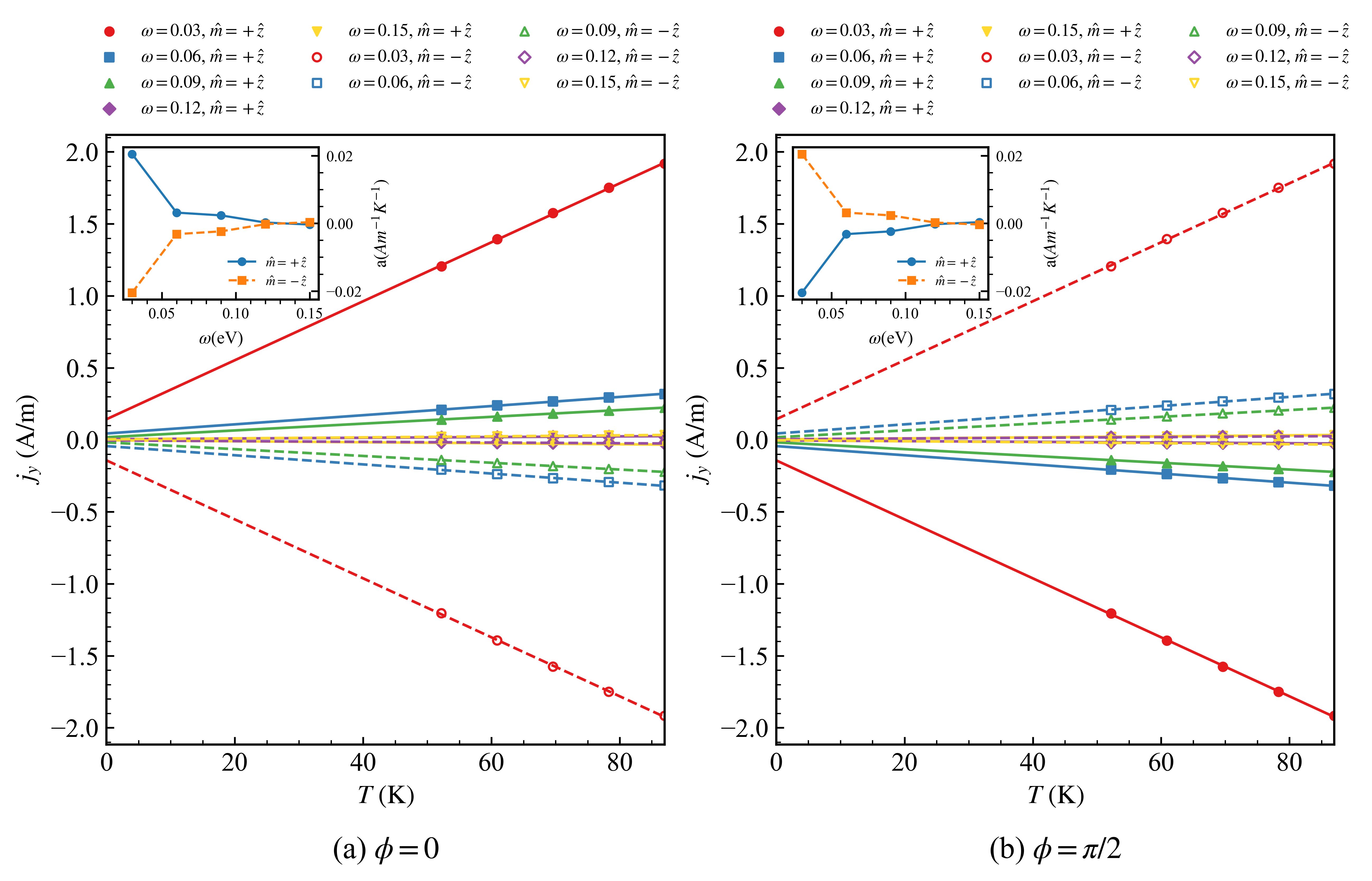}
\caption{\label{fig:jyvsTdiagram}  
  LPGE currents $j_y$ are plotted as a function of temperature $T$ for an applied electric field amplitude $|E|=10^5$ V/m and magnetization strength $J|m| = 0.18$ eV. The polarization angle of the incident electric field and the magnetization direction are indicated in the captions of the individual panels. The currents for $\phi=0$ are calculated from the conductivities using the expression $j_y=\chi^{yxx}|E|^2$ and those for $\phi=\pi/2$ are calculated using $j_y=\chi^{yyy}|E|^2$. The lines represent linear fit of the form $ax+b$. Solid lines represent linear fit for the data corresponding to $\hat{m}=+\hat{z}$, while dotted lines represent data for  $\hat{m}=-\hat{z}$ Insets display the frequency dependence of the extracted slope $a$. The plots demonstrate a linear scaling of the current with temperature whose slopes gradually approach zero}
\end{figure*}

\section{Results and Discussion}
\label{num}

We follow the effective tight-binding model from Refs. \cite{Ozawa2019, Ozawa2024} constructed using the $d_{3z^2 - r^2}$ Co orbital and $p_z$ interlayer Sn orbital. The Sn atom in the Kagome layer and the S atoms were determined to be far from the Fermi level and thus ignored. This Hamiltonian is Fourier transformed in the $x$ and $y$ directions to give a layered Hamiltonian where we identify the $z=0$ Co layer as the top surface layer and the $z=L$ Sn layer as the bottom surface layer where $L$ is the number of layers.   See Appendix. \ref{model} for details about the model and Fig. \ref{fig:structure} for the effective crystal structure. 

\begin{figure*}
\includegraphics[width=1.9\columnwidth]{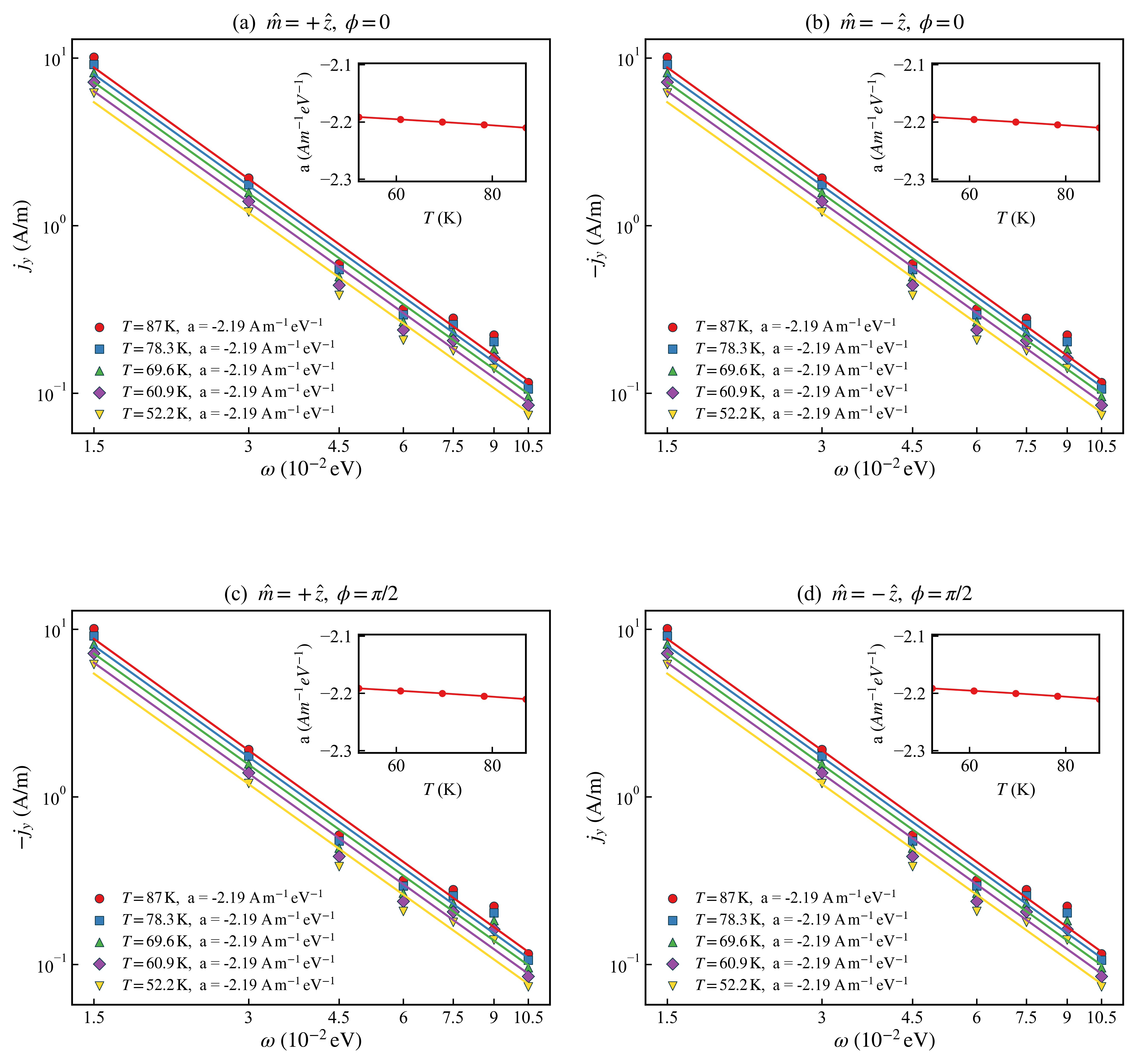}
\caption{\label{fig:jyvswdiagram}  
$j_y$ vs $\omega$ for an applied electric field amplitude $|E| = 10^{5}\,\mathrm{V/m}$ and magnetization strength $J|m| = 0.18\,\mathrm{eV}$. Since the currents in panels (b) and (c) are negative, $-j_y$ is plotted as indicated in the $y$ labels. The polarization angle of the incident electric field and the magnetization direction are indicated in the titles of the individual panels. The currents for $\phi=0$ are calculated from the conductivities using the expression $j_y=\chi^{yxx}|E|^2$ and those for $\phi=\pi/2$ are calculated using $j_y=\chi^{yyy}|E|^2$. Solid lines represent linear fits of the form $\log| j_y| = a \log \omega + b$. Insets display the temperature dependence of the extracted slope $a$. The plot demonstrates a power law scaling of the current with frequency, $j_y \propto\omega^{a}$ where the exponent $a\approx -2.2$ and exhibits weak temperature-dependence.
}
\end{figure*}

To calculate $\chi^{\mu\alpha \beta}$ for magnetization in $+z$ and $-z$ directions, we use a 40-layer system of alternating Co and Sn layers, starting with a Co layer and ending with an Sn layer. The $k$-space integration is performed with $1201 \times 1201$ $k$-points in a 2D Brillouin zone spanned by reciprocal lattice vectors ($ \pi/\sqrt{3} , \pi$)/$a$ and ($\pi/ \sqrt{3} , -\pi$)/$a$ with $a=5.37 \AA$. The band parameters are $t_1 = 0.15$ eV, $J |m| = 0.18$ eV, $t_2=0.09$ eV, $t_z = -0.15$ eV ,  $t_{dp} = 0.27$ eV, $t_{so} = - 0.015$ eV, $\epsilon_p = -1.08$  eV. In addition, we have chosen $\delta=0.03$ eV which corresponds to a lifetime $\tau=22$ fs, and an electric field density of $|E|=10^5$ V/m or light intensity of $\sim 10^7$ W/m$^2$.

For a typical laser spot radius of $\sim 20~\mu$m, this corresponds to an incident optical power of $\sim 10$ mW at the sample.

Using this, we calculate the LPGE ($j_{L}^\mu$) currents for different polarization directions, external frequencies ($\omega$), and temperatures. The calculated currents include both intrinsic and extrinsic contributions and are $O(1)$ A/m.

\begin{figure*}
\centering

\subfloat[$j_x$ vs $\phi$ for $\hat{m}=+\hat{z}$]{
    \includegraphics[width=0.5\columnwidth]{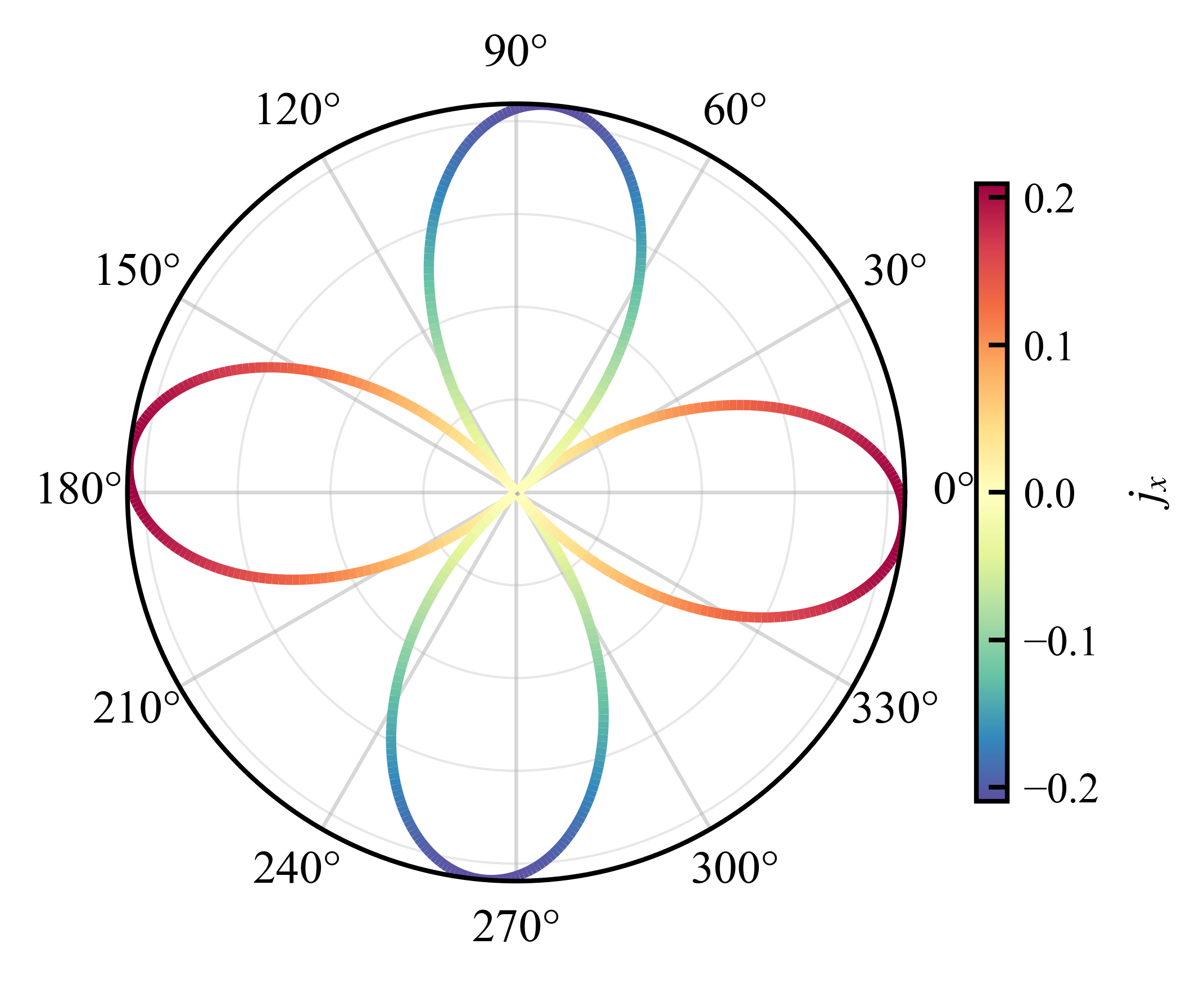}
}
\subfloat[$j_y$ vs $\phi$ for $\hat{m}=+\hat{z}$]{
    \includegraphics[width=0.5\columnwidth]{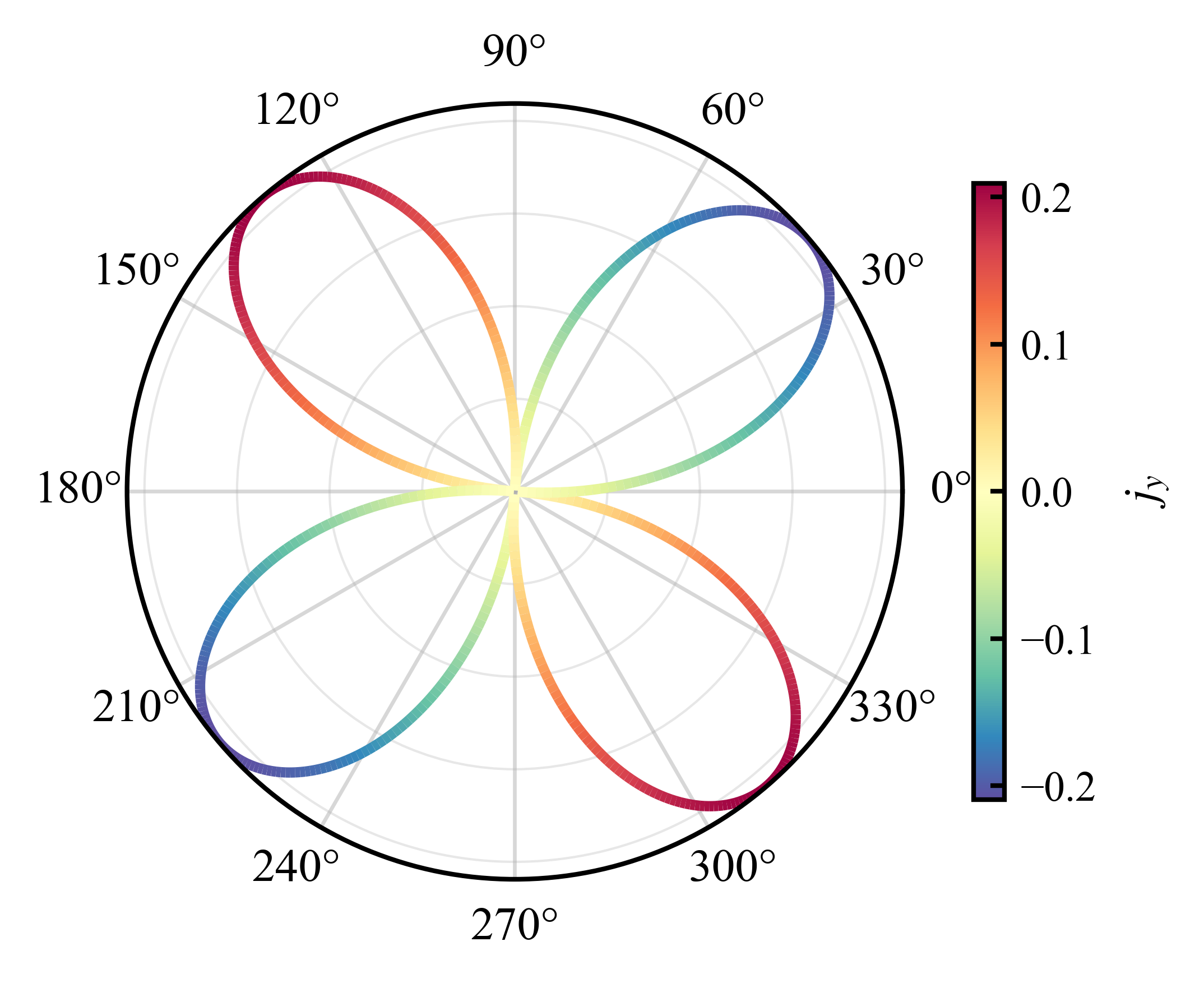}
}
\subfloat[$j_x$ vs $\phi$ for $\hat{m}=-\hat{z}$]{
    \includegraphics[width=0.5\columnwidth]{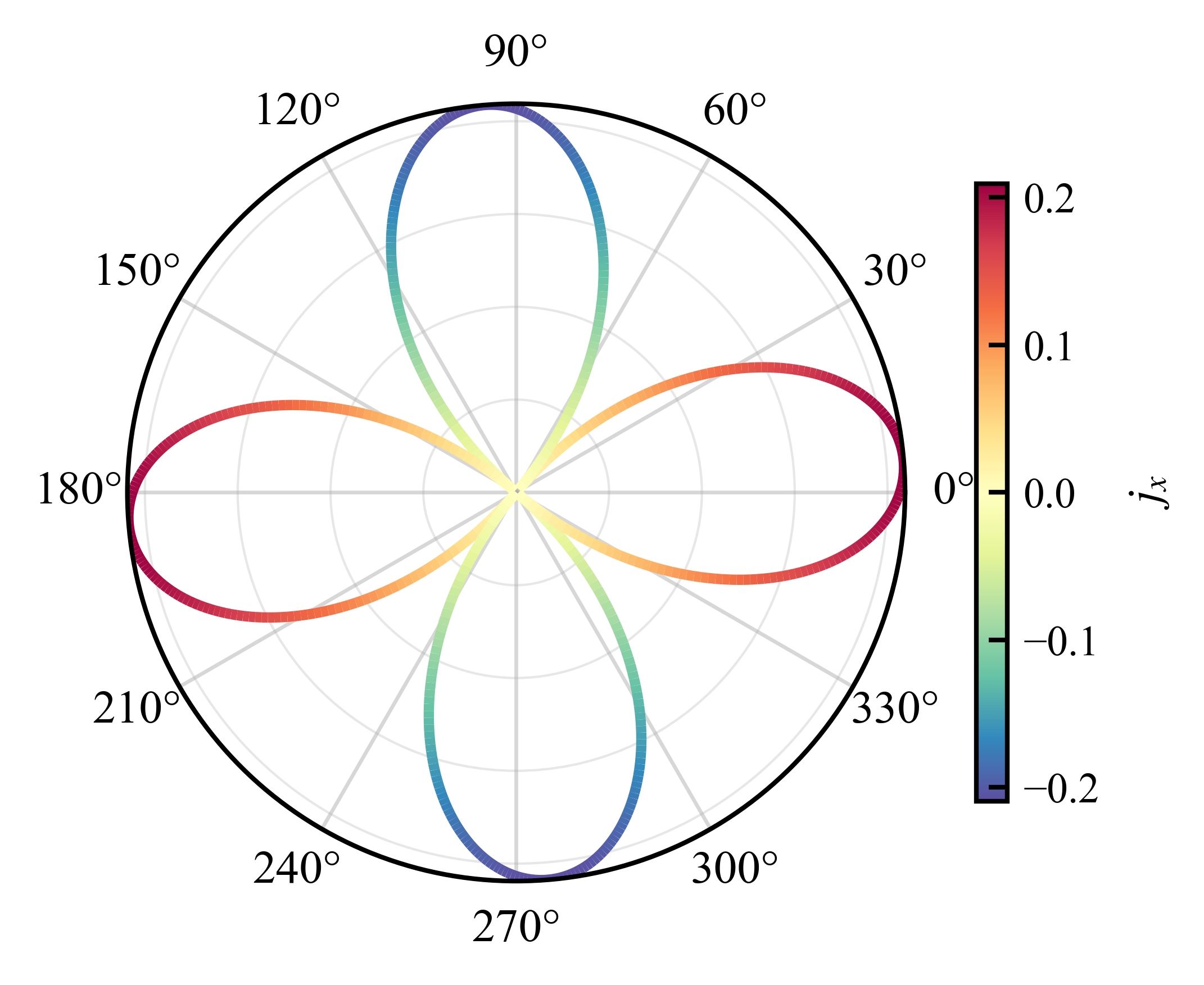}
}
\subfloat[$j_y$ vs $\phi$ for $\hat{m}=-\hat{z}$]{
    \includegraphics[width=0.5\columnwidth]{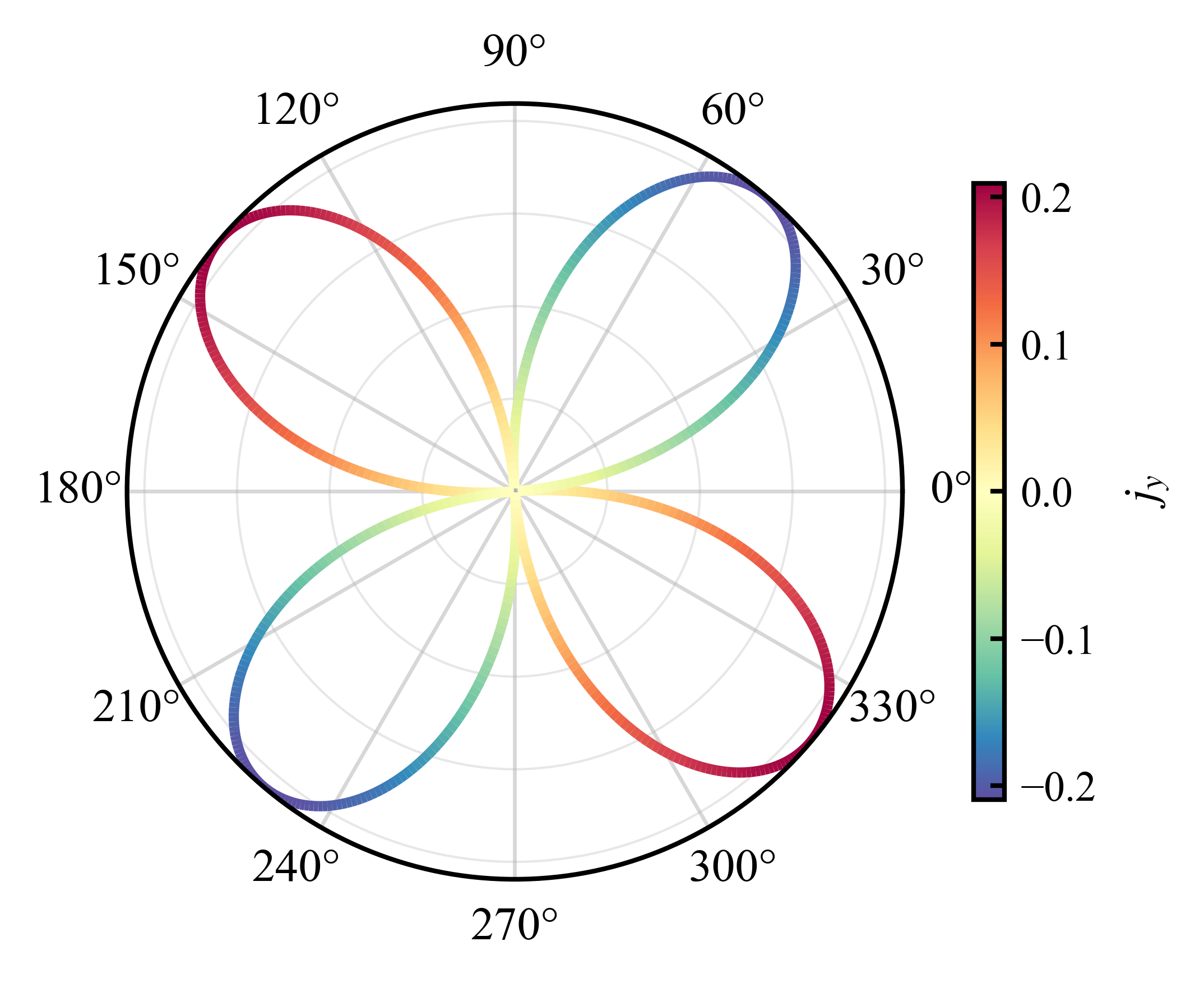}
}

\caption{Polar plots of LPGE currents vs the polarization angle $\phi$ of a linearly polarized electric field for magnetization $\hat{m}=\pm \hat{z}$. Parameters used are $|E|=10^5$ V/m, $J|m|=0.18$ eV, $\hbar\omega=0.15$ eV, and $T=87$ K.}
\label{fig:lpgephi}
\end{figure*}

The plot of the LPGE current $j_y$ as a function of temperature $T$ (Fig.~\ref{fig:jyvsTdiagram}) shows an approximately linear increase of the current with temperature for all frequencies $\omega$. Each curve corresponds to a fixed value of $\omega$, and the temperature dependence becomes weaker at higher frequencies, as reflected in the decreasing slope of the linear fits. For a given polarization angle $\phi$, the cases $\hat{m} = +\hat{z}$ and $\hat{m} = -\hat{z}$ are plotted together. In both $\phi = 0$ and $\phi = \pi/2$ configurations, the current reverses sign upon switching the magnetization direction, consistent with the symmetry of the response. The solid (dashed) lines represent linear fits of the form $j_y = aT + b$ for $\hat{m} = +\hat{z}$ ($\hat{m} = -\hat{z}$). The relatively small values of the slope $a$ indicate a weak temperature dependence of the current. This trend is further illustrated in the inset, which shows the variation of the extracted slope $a$ with frequency $\omega$, confirming that the temperature sensitivity decreases with increasing frequency.

The dependence of the LPGE current on $\hbar\omega$ (Fig.~\ref{fig:jyvswdiagram}) is shown on a log--log scale. For panels (b) and (c), where the current is  negative, the quantity $-j_y$ is plotted in order to represent the data on a logarithmic scale; the corresponding axis labels indicate the same. In the low-frequency regime, the magnitude of the current decreases systematically with increasing $\omega$ across all temperatures. A linear fit in the log--log representation, log $|j_y| = a \log \omega + b$,
yields a slope $a \approx -2.2$, indicating a power-law dependence of the form $|j_y| \propto \omega^{-2.2}$. The extracted exponent exhibits only weak temperature dependence, as shown in the insets, suggesting that the scaling behavior is robust over the temperature range considered.

Figure~\ref{fig:lpgephi} shows polar plots of the LPGE currents $j_x$ and $j_y$ as functions of the polarization angle $\phi$ of a linearly polarized electric field, for magnetization oriented along $\hat{m}=\pm \hat{z}$. Both components exhibit the expected $\pi$-periodic angular dependence. The dominant response occurs away from the symmetry-selected polarization angles and is not, in general, strictly odd under magnetization reversal because the total current contains both even- and odd-in-$m_z$ contributions. However, at special polarization angles, the angular structure isolates tensor components that are odd under $m_z\rightarrow -m_z$. In particular, $j_x$ at $\phi=\pi/4,\ 3\pi/4,\ 5\pi/4,\ 7\pi/4$ and $j_y$ at $\phi=0,\ \pi/2,\ \pi,\ 3\pi/2$ reverse sign exactly upon flipping the magnetization, as predicted by equations \eqref{eq:jx-odd} and \eqref{eq:jy-odd}. Although the current magnitude at these angles is much smaller than the maximum response because the dominant even-in-$m_z$ response is suppressed, these configurations provide the cleanest symmetry fingerprint of the magnetization-odd LPGE.

\section{Summary}
In summary, we investigated the LPGE in $\mathrm{Co}_3\mathrm{Sn}_2\mathrm{S}_2$ using the diagrammatic perturbation approach based on an effective tight-binding model. We find a large effect with strong dependence on magnetization direction, polarization angle, temperature, and driving frequency. The current shows an approximately linear temperature dependence, and follows a low-frequency power-law scaling of the form $|j_y| \propto \omega^{-2.2}$ with only weak temperature dependence of the scaling exponent. This observation, along with the fact that the LPGE must vanish in the bulk due to centrosymmetry, indicates that the large size of the response may be due to the significant density of states from the surface FAs. The angular dependence of the photocurrent further reveals special high-symmetry polarization angles at which the photocurrent exhibits clear odd behavior under magnetization reversal. These results establish the centrosymmetric, magnetic WSM $\mathrm{Co}_3\mathrm{Sn}_2\mathrm{S}_2$ as a promising platform for studying symmetry-controlled nonlinear optoelectronic phenomena in magnetic topological materials.
\newline
\newline
\begin{acknowledgments}
A.N., K.C. and P.H. were supported by the National Science Foundation grant no. DMR 2047193. N.S. acknowledges financial support from the Prime Minister’s Research Fellows (PMRF) scheme offered by the Ministry of Education, Government of India. N.S. and H.K. thank the National Supercomputing Mission (NSM) for providing computing resources of “PARAM Rudra” at IITB, implemented by C-DAC and supported by the Ministry of Electronics and Information Technology (MeitY) and
the Department of Science and Technology (DST), India. A.N. would like to thank Siddharth Mansingh for useful discussions that aided numerical computations.  
\end{acknowledgments}

\begin{widetext}

\appendix
\section{Effective Tight-Binding Model for $\text{Co}_3\text{Sn}_2\text{S}_2$}
\label{model}
The tight binding Hamiltonian for $\text{Co}_3\text{Sn}_2\text{S}_2$ obtained from Ozawa and Nomura(2019) \cite{Ozawa2019, Ozawa2024} is written as 
\begin{equation}
    H = H_{hop} + H_{so} + H_{exc}
\end{equation}
$H_{hop}$ contains the hopping between nearest-neighbor and next nearest-neighbor Co-Co terms in the Kagome layer, interlayer nearest-neighbor Sn-Co terms, and onsite energies for Sn and Co.  $H_{so}$ is the spin-orbit coupling,  and  $H_{exc}$ is the coupling term between spins and magnetic moment. In momentum space, this looks like:

\begin{align}
    \hat{H}_{\boldsymbol{k}} =& \sum_{\boldsymbol{k}ss'}\epsilon_{p}p_{\boldsymbol{k}s}^{\dagger}p_{\boldsymbol{k}s}-Jd_{\boldsymbol{k}As}^{\dagger} (\textbf{m}\cdot\boldsymbol{\sigma})_{ss'} \ d_{\boldsymbol{k}As'}-Jd_{\boldsymbol{k}Bs}^{\dagger}(\textbf{m}\cdot\boldsymbol{\sigma})_{ss'}\ d_{\boldsymbol{k}Bs'}-J d_{\boldsymbol{k}Cs}^{\dagger}(\textbf{m}\cdot\boldsymbol{\sigma})_{ss'}\ d_{\boldsymbol{k}Cs'} \nonumber \\ 
	&-2it_{dp}\sin(k_{1a})d_{\boldsymbol{k}As}^{\dagger}p_{\boldsymbol{k}s}-2it_{dp}\sin(k_{2a})d_{\boldsymbol{k}Bs}^{\dagger}p_{\boldsymbol{k}s}-2it_{dp}\sin(k_{3a})d_{\boldsymbol{k}Cs}^{\dagger}p_{\boldsymbol{k}s} \nonumber \\ 
    & +  [-2t_{1}\cos(k_{1b})-2t_{2}\cos(k_{1d})-2t_{z}\cos(k_{1c})+2it_{so}\cos(k_{1d})\sigma_{z}]d_{\boldsymbol{k}As}^{\dagger}d_{\boldsymbol{k}Bs}
\nonumber \\ 
    &	+[-2t_{1}\cos(k_{3b})-2t_{2}\cos(k_{3d})-2t_{z}\cos(k_{3c})-2it_{so}\cos(k_{3d})\sigma_{z}]d_{\boldsymbol{k}As}^{\dagger}d_{\boldsymbol{k}Cs}
\nonumber \\ 
    &	+[-2t_{1}\cos(k_{2b})-2t_{2}\cos(k_{2d})-2t_{z}\cos(k_{2c})+2it_{so}\cos(k_{2d})\sigma_{z}]d_{\boldsymbol{k}Bs}^{\dagger}d_{\boldsymbol{k}Cs}+h.c.
\end{align}

Here, $k_{ia}=\boldsymbol{k}\cdot \textbf{a}_i$, $k_{ib}=\boldsymbol{k}\cdot \textbf{b}_i$, $k_{ic}=\boldsymbol{k}\cdot \textbf{c}_i$ and $k_{id}=\boldsymbol{k}\cdot \textbf{d}_i$. The vectors $\textbf{a}_i$, $\textbf{b}_i$, $\textbf{c}_i$, $\textbf{d}_i$ are defined in references \cite{Ozawa2019, Ozawa2024}. $t_1$ is the probability amplitude for intralayer nearest-neighbor Co-Co hopping along the vector $\textbf{b}_i$, $t_2$ for next nearest-neighbor Co-Co hopping along the vector $\textbf{d}_i$ and $t_z$ for interlayer nearest-neighbor Co hopping along the vector $\textbf{c}_i$. $\epsilon_p$ is the energy difference between $p$ and $d$ orbitals and $t_{dp}$ is the nearest-neighbor Co-Sn hopping along the vector $\textbf{a}_i$. $J$ is the exchange coupling constant, $\boldsymbol{m}$ denotes the magnetization spin direction and $\boldsymbol{\sigma}=(\sigma_x,\sigma_y,\sigma_z)$ denote the Pauli matrices acting in spin-space. $t_{so}$ is the spin-orbit coupling amplitude. $d^\dagger_{\boldsymbol{k}is}$  is the creation operator for the  Co sites and $p^\dagger_{\boldsymbol{k}s}$ is the creation operator for the  Sn sites. $i=A,B,C$ refers to the 3 Co sites and $s=\uparrow,\downarrow$ refers to the spin direction.

This Hamiltonian is partially Fourier transformed only in the z-direction to give a slab Hamiltonian which looks like:

\begin{align}
    \hat{H}_{k_x k_y} =      \boldsymbol{c}^\dagger_{k_x k_y }\begin{pmatrix}
\mathcal{H}_{k_x k_y} & h_{k_x k_y}         &       0     &   \cdots &0     \\
h_{k_x k_y}^\dagger  & \mathcal{H}_{k_xk_y} &  h_{k_x k_y}&   &         \\
       \vdots        &                 & \ddots     &           \\
        0             &             & & h_{k_x,k_y}^\dagger  & \mathcal{H}_{k_x,k_y} 
\end{pmatrix}
\boldsymbol{c}_{k_x k_y} 
\end{align}

$\hat{H}$ is now an $8L\times8L$ matrix, where $L$ is the number of unit cells forming the slab. Each unit cell consists of a Co-Sn bilayer. Each $\mathcal{H}_{k_x k_y}$ is an $8 \times 8$ block matrix, $h_{k_x k_y}$ is the interlayer hopping block and  
\begin{equation}
\boldsymbol{c}_{k_x k_y } = \begin{pmatrix}
d_{k_x k_y 0  A   \uparrow} & d_{k_x k_y 0   A  \downarrow} & d_{k_x k_y 0  B  \uparrow} & d_{k_x k_y 0   B  \downarrow} & d_{k_x k_y 0  C \uparrow} & d_{k_x k_y  0  C \downarrow} & p_{k_x k_y 0   \uparrow} & p_{k_x k_y 0  \downarrow} & d_{k_x k_y 1  A  \uparrow} \cdots \\
\end{pmatrix}
\end{equation}

$\mathcal{H}_{k_x k_y}$ is given as 
\begin{align}
\mathcal{H}_{k_x k_y} = \begin{pmatrix}
-J(\textbf{m}\cdot\boldsymbol{\sigma}) & -2(t_{1}\cos k_{1b} + t\cos k_{1d}) & -2(t_{1}\cos k_{3b} + t^*\cos k_{3d}) & -t_{dp}e^{ik_{1a}} \\
-2(t_{1}\cos k_{2b} + t^*\cos k_{2d}) & -J(\textbf{m}\cdot\boldsymbol{\sigma}) & -2(t_{1}\cos k_{2b} + t\cos k_{2d}) & -t_{dp}e^{ik_{2a}} \\
-2(t_{1}\cos k_{3b} + t\cos k_{3d}) & -2(t_{1}\cos k_{2b} + t^*\cos k_{2d}) & -J(\textbf{m}\cdot\boldsymbol{\sigma}) & -t_{dp}e^{ik_{3a}} \\
-t_{dp}e^{-ik_{1a}} & -t_{dp}e^{-ik_{2a}} & -t_{dp}e^{-ik_{3a}} & \epsilon_{p}
\end{pmatrix}   
\end{align}

and

\begin{align}
    h_{k_x k_y} = \begin{pmatrix}
0  & -t_{z}e^{ik_{1c}} & -t_{z}e^{ik_{3c}} & 0 \\
-t_{z}e^{ik_{1c}}  & 0 & -t_{z}e^{ik_{2c}} & 0 \\
 -t_{z}e^{ik_{3c}} & -t_{z}e^{ik_{2c}} & 0 & 0 \\
t_{dp}e^{ik_{1a}} & t_{dp}e^{ik_{2a}} & t_{dp}e^{ik_{3a}} & 0
\end{pmatrix} 
\end{align}

where t is a complex number defined as $t=t_2-it_{so}$.

\end{widetext}

\bibliography{library}

\end{document}